\documentclass[runningheads]{llncs}
\usepackage[T1]{fontenc}
\usepackage{graphicx}
\usepackage{changepage}
\usepackage{paralist}
\usepackage{multirow}
\usepackage{amsmath}
\usepackage{txfonts}
\usepackage{amsfonts}
\usepackage{mathabx}
\usepackage{makecell}
\usepackage{booktabs}
\usepackage{CJKutf8}

\usepackage{hyperref}
\hypersetup{
	colorlinks=true,
	linkcolor=cyan,
	citecolor=green,
}
\usepackage{color}

\makeatletter
\newcommand\footnoteref[1]{\protected@xdef\@thefnmark{\ref{#1}}\@footnotemark}
\makeatother

\begin{document}
\begin{CJK}{UTF8}{gkai}

\title{KunquDB: An Attempt for Speaker Verification in the Chinese Opera Scenario}

\author{
Huali Zhou\inst{1,2} \and
Yuke Lin\inst{1,2} \and
Dong Liu\inst{2} \and
Ming Li\inst{1,2}\thanks{Corresponding author. E-mail: \href{mailto:ming.li369@dukekunshan.edu.cn}{ming.li369@duke.edu}}
}
\authorrunning{H. Zhou et al.}

\institute{
School of Computer Science, Wuhan University, Wuhan, China \and
Suzhou Municipal Key Laboratory of Multimodal Intelligent Systems,\\Data Science Research Center, Duke Kunshan University, Kunshan, China\\
\email{hualizhou@whu.edu.cn, \{yuke.lin,dong.liu,ming.li369\}@dukekunshan.edu.cn}
}

\maketitle

\begin{abstract}
This work aims to promote Chinese opera research in both musical and speech domains, with a primary focus on overcoming the data limitations. We introduce KunquDB, \url{https://hualizhou167.github.io/KunquDB}, a relatively large-scale, well-annotated audio-visual dataset comprising $339$ speakers and $128$ hours of content. Originating from the Kunqu Opera Art Canon (\textit{Kunqu yishu dadian}), KunquDB is meticulously structured by dialogue lines, providing explicit annotations including character names, speaker names, gender information, vocal manner classifications, and accompanied by preliminary text transcriptions. KunquDB provides a versatile foundation for role-centric acoustic studies and advancements in speech-related research, including Automatic Speaker Verification (ASV). Beyond enriching opera research, this dataset bridges the gap between artistic expression and technological innovation. Pioneering the exploration of ASV in Chinese opera, we construct four test trials considering two distinct vocal manners in opera voices: stage speech (\textbf{\textit{ST}}) and singing (\textbf{\textit{S}}). Implementing domain adaptation methods effectively mitigates domain mismatches induced by these vocal manner variations while there is still room for further improvement as a benchmark.

\end{abstract}

\section{Introduction}
Chinese opera, or \textit{Xiqu}, is a distinguishable and traditional art form that has gained worldwide recognition. Kunqu Opera, Beijing Opera, and Cantonese Opera have been proclaimed World Intangible Cultural Heritage, highlighting their exceptional artistic contributions and rich cultural heritage. Chinese opera is a confluence of song, speech, mime, dance, and acrobatics, bound together by theatrical conventions that differ significantly from Western opera~\cite{lin2022modernising}. 

As a distinctive form of performing arts, Chinese opera diverges from conventional speech and typical singing. In the realm of speech research, opera provides a distinctive experimental ground, given its intricate fusion of speech, music, and theatrical elements. The multifaceted acoustic expressions within opera voices create an exceptional context for in-depth exploration in speech research. Regardless, previous research on Chinese opera has predominantly stemmed from musical and literary perspectives, relying on traditional methodologies rather than integrating state-of-the-art technical tools. The absence of automated deep-learning tools has led to a heavy reliance on manual data pipelines for collecting and annotating Chinese opera datasets. Consequently, existing opera datasets~\cite{caro2014creating,islam2015chinese,gong2019jingju,chen2022sustainable} face limitations in terms of scale and annotation richness, typically covering only a few hours~\cite{islam2015chinese,chen2022sustainable} and providing genre information exclusively~\cite{chen2022sustainable}. In contrast to the comprehensive annotations provided in speech and singing datasets, which include speaker labels, text transcriptions, phoneme-level durations, and pitch information, existing Chinese opera datasets lack comparable richness.

The scarcity of detailed annotations poses a significant obstacle for numerous research tasks on opera data. This obstacle is particularly pronounced for tasks requiring comprehensive annotations, including automatic speaker recognition for speaker label prediction, Automatic Speech Recognition (ASR) for text transcription retrieval, speaker diarization for role detection, as well as speech and singing voice synthesis. Meanwhile, speech-related research predominantly focuses on conventional speech. Existing open-source models designed for various speech tasks, such as speaker diarization, exhibit inadequate robustness when applied to opera data. The complex acoustic characteristics in opera voices provide a diverse testing ground for evaluating the robustness of speech models. The absence of automated tools further obstructs large-scale data collection and cleaning, restricting access to diverse and abundant datasets. This dilemma creates a cycle that impedes progress in data availability, hindering the development of advanced tools for digitizing opera research. 

In response to the challenges posed by insufficient data and limitations of existing models in the field of Chinese opera, our primary objective is to create a symbiotic relationship between data and models. To achieve this, we present a comprehensive and publicly accessible audio-visual dataset characterized by its richness and scale. This resource is designed to lay the groundwork for developing specialized automated tools applicable to Chinese opera, thereby facilitating advancements in the study of this art form. Narrowing down from the landscape of Chinese traditional opera, we focus on one exquisite domain, Kunqu Opera. Reputed as the mother of Chinese operas, Kunqu Opera boasts a history spanning over 600 years~\cite{dong2014loudness}, giving rise to numerous operas, including Beijing Opera. In alignment with \cite{caro2014creating}, we selectively choose classic and authoritative audio-visual materials sourced from the Kunqu Opera Art Canon (\textit{Kunqu yishu dadian})\footnote{\label{publisher}\textbf{Note}: After purchasing the book, we negotiated with the publisher and secured their authorization for its utilization in Kunqu Opera research. The publisher explicitly stated that the book's digital resource can be employed solely for scholarly or research endeavors upon the approval of the publisher. It may not be illegally disseminated or used for commercial purposes.}~\cite{wang2016kunquyishudadian} to ensure both quantity and quality. The source video undergoes sentence-level segmentation, generating preliminary text transcriptions. Subsequently, we proceed with speaker annotation and explicitly categorize each utterance as either stage speech (\textbf{\textit{ST}}) or singing (\textbf{\textit{S}}) based on vocal manner. Ultimately, KunquDB\footnote{\label{url-kunqudb}\url{https://hualizhou167.github.io/KunquDB}}, the curated audio-visual dataset comprises 339 performers, totals approximately 128 hours, with stage speech and singing voices each constituting about half of the dataset. As an audio-visual dataset, it is applicable in various scenarios, including ASV, ASR, speaker diarization, singing voice synthesis, person re-identification and multi-modal understanding.

Building on KunquDB, we investigate automatic speaker verification within the Kunqu Opera context. We aim to provide insights that enhance subsequent synthesis efforts, accommodating variations in role types and vocal manners. Speaker verification in Kunqu Opera bears similarities to the task in \cite{brown2021playing}, involving speech from interviews (typically calm and quiet) and speech from movies (with varying emotion and background noise) across different domains. This yields a cross-domain speaker verification challenge induced by vocal manner, an internal factor of the speaker~\cite{nagrani2017voxceleb}. To tackle cross-domain issues, we implement domain adversarial training, leveraging domain prediction to obtain speaker-discriminative and domain-invariant representations. Furthermore, we employ the batchwise Siamese training strategy to maintain consistency across different vocal manners for the same speaker. Experimental results validate the efficacy of the domain adaptation methods.

Our main contributions are summarized as follows:
\begin{asparaitem}
    \item We curate KunquDB\footnoteref{url-kunqudb}, a comprehensive audio-visual dataset specifically tailored for Kunqu Opera. Its large scale effectively mitigates data shortages and fosters a positive feedback loop between data and tool models.
    \item To the best of our knowledge, we are the first to explore ASV within Chinese opera, addressing mismatches across stage speech and singing. The implementation of domain adaptation methods sets a benchmark for future research.
\end{asparaitem}

\vspace{-5pt}
\section{Vocal Distinctions in Chinese Opera Versus Speech}
\vspace{-5pt}
The aural aspect of Chinese traditional opera significantly differs from ordinary spoken and contemporary singing, including textual structure, pronunciation, intonation, vocal manner, and overall expressive forms. 
\begin{inparaenum}[(1)]
    \item The textual dimension of Chinese traditional opera involves two types: song lyrics (\textit{changci}) for expressing emotions and stage speech (\textit{nianbai}) for advancing the narrative~\cite{wichmann1991listening}. Within the text, two linguistic levels emerge: classical Chinese (\textit{wenyan wen}), an archaic written language, and vernacular (\textit{baihua}), which includes standard spoken Mandarin or regional dialects with distinct phonetic variations.
    \item From a melodic perspective, Chinese opera draw its musical compositions from a pre-existing repertoire of tunes. Unlike Western opera, where a designated "composer" is assigned, in Chinese opera, the scriptwriter selects tunes deemed suitable for the dramatic context from the repertoire and crafts the accompanying text. Musical notation is absent; instead, the script specifies tunes by name, with the text intended to be sung accordingly~\cite{yung1983creative}. Notably, stage speech and singing exhibit considerably higher Equivalent Sound Levels (Leq) compared to regular speech \cite{dong2014loudness}.
    \item From a vocalization standpoint, Chinese opera utilizes two vocal techniques: "false-voice" (\textit{jiasangzi}), executed in falsetto, and "true-voice" (\textit{zhensangzi}), produced by vocal cord vibration. Falsetto serves various purposes. Firstly, male actors, exemplified by renowned figures like Mei Lanfang, portray female characters in Chinese opera, employing falsetto to imitate the female voice~\cite{jinpei1989xipi}. Secondly, falsetto is believed to ideally produce essential, extended sounds pronounced with a nearly closed mouth~\cite{wichmann1991listening}. 
\end{inparaenum}

\begin{figure}[t]\footnotesize
  \centering
  \includegraphics[width=\linewidth]{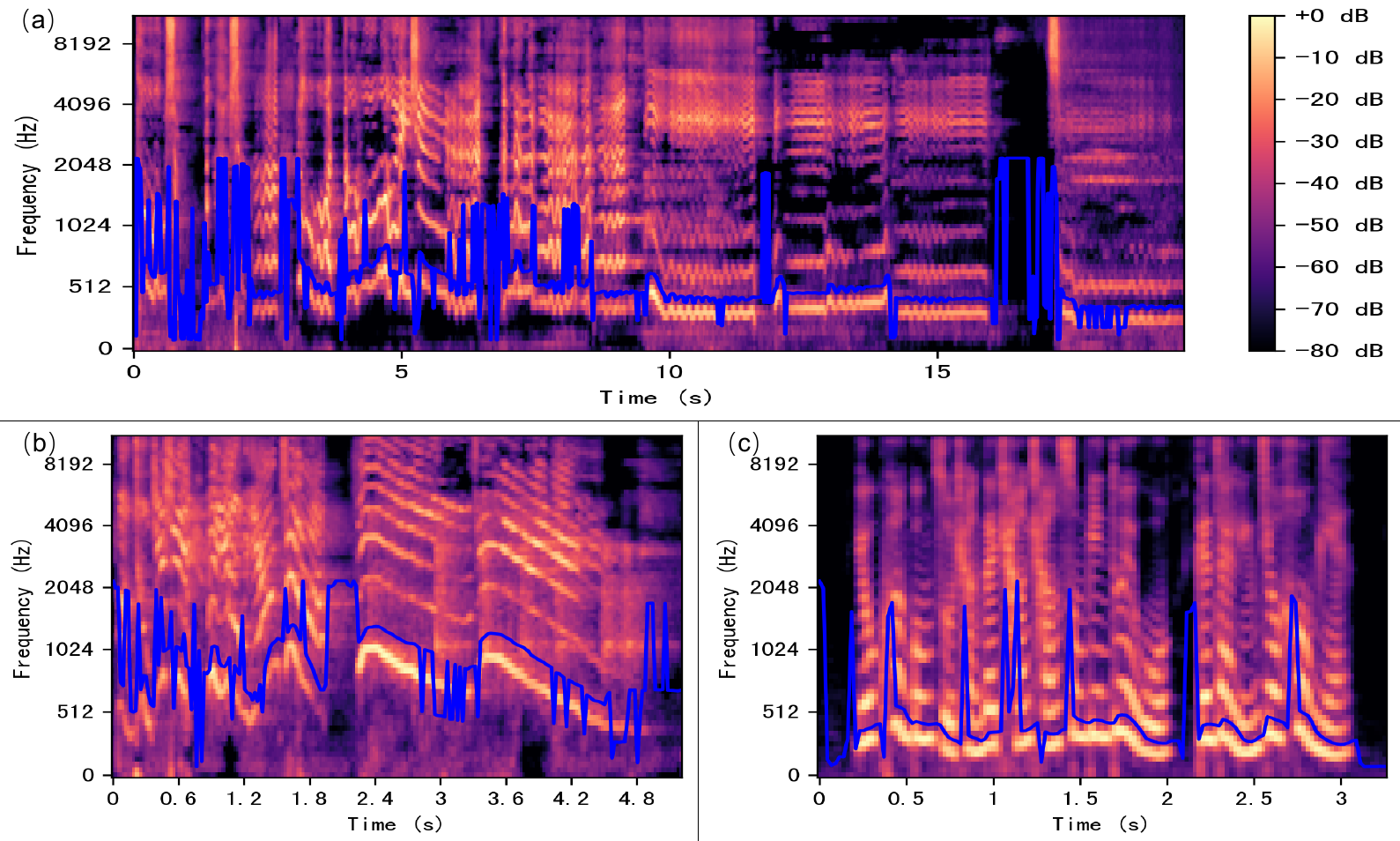}
  \caption{Mel spectrograms with overlaid pitch contours for singing (\textbf{a}), stage speech (\textbf{b}), and regular speech (\textbf{c}).}
  \label{fig-mels}
\end{figure}

Fig.~\ref{fig-mels} displays Mel spectrograms with overlaid pitch contours for randomly selected utterances representing singing, stage speech, and regular speech (from an external speech dataset). Singing and stage speech consistently exhibit higher frequency compared to regular speech. Moreover, singing showcases more dynamic pitch variation than stage speech, highlighting two distinct acoustic characteristics in Chinese opera.

\vspace{-5pt}
\section{Related Works}
\subsection{Chinese Opera}
\vspace{-3pt}
Open-source datasets for Chinese opera remain limited. While \cite{islam2015chinese} and \cite{chen2022sustainable} propose datasets for opera genre and Cantonese singing genre classification, respectively, these datasets are not publicly accessible. Due to the lack of publicly available corpora, \cite{zhang2021chinese,li2023joint,yao2024analysis} targeting opera genre classification, collect data individually for personalized experiments. Typically, these datasets consist of individual instances structured as audio files paired with corresponding Chinese opera genre labels. On the other hand, open-access datasets have driven advancements in academic research. For example, the CompMusic Beijing Opera corpus proposed in \cite{serra2014creating} aids \cite{srinivasamurthy2014transcription} in acquiring Beijing Opera percussion patterns for transcription and recognition. Similarly, the unaccompanied singing data released by \cite{black2014automatic} provides the foundation for \cite{yang2015vibrato} to analyze pitch histograms and vibrato statistics in Beijing Opera singing. 

Due to the lack of automatic tools for data collection and cleaning, existing opera datasets~\cite{caro2014creating,islam2015chinese,gong2019jingju,chen2022sustainable} are limited in scale and annotation richness. Typically spanning only a few hours~\cite{islam2015chinese,chen2022sustainable} and offering plain labels~\cite{chen2022sustainable}, they are insufficient for downstream recognition tasks like ASV and singing speech recognition in the field of Chinese opera. In the intersection of opera and speech-related research, most efforts are focused on synthesis, with reliance on the only publicly accessible, well-annotated yet small-scale dataset, "Jingju a cappella singing"~\cite{gong2019jingju}. It serves as the basis for subsequent neural network-based opera synthesis by~\cite{wu2019synthesising,wu2020peking,zhou2023high,peng2023singing}. While \cite{wu2019synthesising} and \cite{wu2020peking} pioneer neural network-based synthesis using the DurIAN~\cite{yu2019durian} framework, \cite{zhou2023high} introduces OperaSinger, based on the FastSpeech2~\cite{ren2020fastspeech} framework, exploring novel data augmentations within this small-scale dataset~\cite{gong2019jingju}. In a related vein, \cite{peng2023singing} attempts to transfer popular singers' timbre to Chinese opera using the VITS~\cite{kim2021conditional} model with the same dataset~\cite{gong2019jingju}. 

\vspace{-5pt}
\subsection{Automatic Speaker Verification}
\vspace{-5pt}
\label{subsec-asv}
Automatic Speaker Verification (ASV) aims to verify whether a given utterance (test utterance) matches the claimed identity by comparing it with the speaker's known utterance (enrollment utterance).  The rise of DNNs in recent years has triggered the evolution of ASV systems from traditional probabilistic models~\cite{reynolds2000speaker,dehak2010front} to deep embedding models~\cite{snyder2018x,desplanques2020ecapa}. A typical DNN-based ASV architecture consists of key components, including:
\begin{inparaenum}[(i)]
\item \textbf{neural network backbone}~\cite{wang2023wespeaker,desplanques2020ecapa,han2022local} (encoder),
\item \textbf{pooling layer}~\cite{wang2021revisiting,snyder2017deep,okabe2018attentive} for temporal aggregation,
\item \textbf{loss function}~\cite{wang2018additive,deng2019arcface} for training optimization,
\item \textbf{scoring strategy}~\cite{prince2007probabilistic} for assessing similarity between embeddings.
\end{inparaenum}

The neural network backbone, as the encoder, extracts frame-level features from the input utterance. This backbone has evolved from architectures like 2D Convolutional Neural Networks (CNNs)~\cite{wang2023wespeaker}, Time Delay Neural Networks (TDNNs)~\cite{desplanques2020ecapa}, and Transformers~\cite{han2022local}. Currently, 2D CNNs with ResNet~\cite{he2016deep} are the most widely adopted. The pooling layer aggregates frame-level features into a fixed-length, utterance-level representation, which is then projected linearly to generate the speaker embedding. Common temporal aggregation techniques include average pooling~\cite{wang2021revisiting}, statistical pooling~\cite{snyder2017deep} and attentive pooling~\cite{okabe2018attentive}. The loss function is the optimized objective during training, such as the Additive Margin Softmax (AM-Softmax)~\cite{wang2018additive} and ArcFace~\cite{deng2019arcface}. The scoring strategy, or the back-end model, measures the similarity between enrollment and test utterance embeddings for verification. Typically, cosine similarity or Probabilistic Linear Discriminant Analysis (PLDA)~\cite{prince2007probabilistic} are utilized.

Despite significant progress in ASV, speaker embeddings' robustness falters with domain shifts, facing challenges from real-world variations~\cite{nagrani2017voxceleb}, resulting in performance degradation. For extrinsic factors, \cite{cai2020within} and \cite{qin2022robust} target noise and far-field conditions for more robust voiceprint representation. Addressing internal factors, \cite{qin2022cross} and \cite{brown2021playing} investigate cross-age and diverse emotional scenes, respectively, to further enhance the robustness.

\vspace{-5pt}
\section{KunquDB Dataset\protect\footnoteref{url-kunqudb}}
\vspace{-5pt}
To obtain authentic singing data for Kunqu Opera and ensure an ample dataset, we leverage audio-visual materials from the authoritative Kunqu Opera Art Canon (\textit{Kunqu yishu dadian})\footnoteref{publisher}~\cite{wang2016kunquyishudadian} as reliable sources. The source videos in this collection~\cite{wang2016kunquyishudadian} contain credits, dialogue lines, and information about vocal manner categories (\textbf{\textit{ST}} or \textbf{\textit{S}}), all of which are hard-coded directly or indirectly.

\subsection{Overall: QAs about KunquDB}
\subsubsection{What is KunquDB?}
KunquDB is a Kunqu Opera audio-visual dataset derived from videos featuring manual annotations for opera character names, speaker identity (ID) labels, gender information, singing/stage speech category labels, and preliminary text transcriptions.

\vspace{-10pt}
\subsubsection{Why is Manual Labeling Required?}
Due to the nature of Kun Opera performances, where the entire stage is often captured rather than close-ups of characters, human faces occupy limited space in the frame. Moreover, performers' heavy makeup and theatrical costumes further obscure facial features, particularly the waist-length beards (\textit{rankou})~\cite{wichmann1991listening} worn by male characters typically completely conceal their mouths. Consequently, conventional pipelines, as used in ~\cite{nagrani2017voxceleb,lin2024voxblink2}, involving face detection, tracking, verification, and audio-video synchronization for mouth movement and speech, are unsuitable for these opera videos.

\vspace{-10pt}
\subsubsection{How to Get KunquDB?}
The book~\cite{wang2016kunquyishudadian} purchase grants access to the digital source video data in a supplementary disc. It is the user's responsibility to get the approval from the publisher to conduct research for non-commercial purposes. We provide annotated data, including segment start and end timestamps, along with associated information, such as character names, speaker names, and preliminary text transcriptions. The open-source annotations and processing scripts can be accessed and downloaded online\footnoteref{url-kunqudb}.

\subsection{Data Collection Pipeline}
\subsubsection{Step 1: Video Segmentation}
We utilize VideoSubFinder\footnote{\url{https://sourceforge.net/projects/videosubfinder}}, in conjunction with PaddleOCR\footnote{\url{https://github.com/PaddlePaddle/PaddleOCR}} to extract hardcoded subtitles from source videos, yielding timestamps for each dialogue line and corresponding text transcriptions. Using ffmpeg\footnote{\label{ffmpeg}\url{https://ffmpeg.org}}, we then segment the videos into clips based on the acquired timestamps, resulting in individual video clips for each dialogue line.

\vspace{-10pt}
\subsubsection{Step 2: Manual Labeling}
The manual annotation process includes categorizing vocal manner and active speaker annotations. Vocal manner annotation is straightforward, with stage speech and singing categorized based on the font style in the original video subtitles. Active speaker annotation is detailed below and is divided into
\begin{inparaenum}[(i)]
  \item discriminative speaker tag,
  \item tag-character annotation, and
  \item character-performer mapping based on each play.
\end{inparaenum}
Eventually, the dataset is structured per dialogue line, encompassing all lines delivered by each performer across different plays.
\begin{compactenum}[i]
  \item \label{item-tag_spk} We recruit $20$ graduate students to assign active speaker tags, each annotating an average of $8.5$ hours of source videos. Participants use XnView MP\footnote{\url{https://www.xnview.com/en/xnviewmp}} software to tag active speakers for each line while watching the complete source video. They adhere to a naming format like \textit{spk\_01} to ensure consistency and avoid repetition within each play. Overlapping speech segments are instructed to be discarded.
  \item Match the active speaker tags akin to \textit{spk\_01} obtained in \ref{item-tag_spk} with the corresponding characters in each play.
  \item Extract character-performer mapping by digitizing the embedded credits in source videos.
\end{compactenum}

\vspace{-10pt}
\subsubsection{Step 3: Extract Audio from Video}
Initially, we use ffmpeg\footnoteref{ffmpeg} to extract 48kHz stereo audio from video segments, then Spleeter~\cite{hennequin2020spleeter} isolates background music, and finally ffmpeg\footnoteref{ffmpeg} downsamples the audio to mono-channel at 16kHz.

\vspace{-10pt}
\subsubsection{Step 4: Assessment and Recheck}
We extract speaker embeddings for individual utterances, using WeSpeaker's~\cite{wang2023wespeaker} ResNet34-based model pretrained on Cn-Celeb~\cite{fan2020cn}. Then, we compute average embeddings for each speaker in each category and assess the cosine similarity between each utterance's embedding and the corresponding average. Utterances with a similarity score below the threshold of $0.4$ undergo manual review.

\vspace{-5pt}
\subsection{Dataset Statistics}
\begin{figure}[t]\footnotesize
  \centering
  \includegraphics[width=\linewidth]{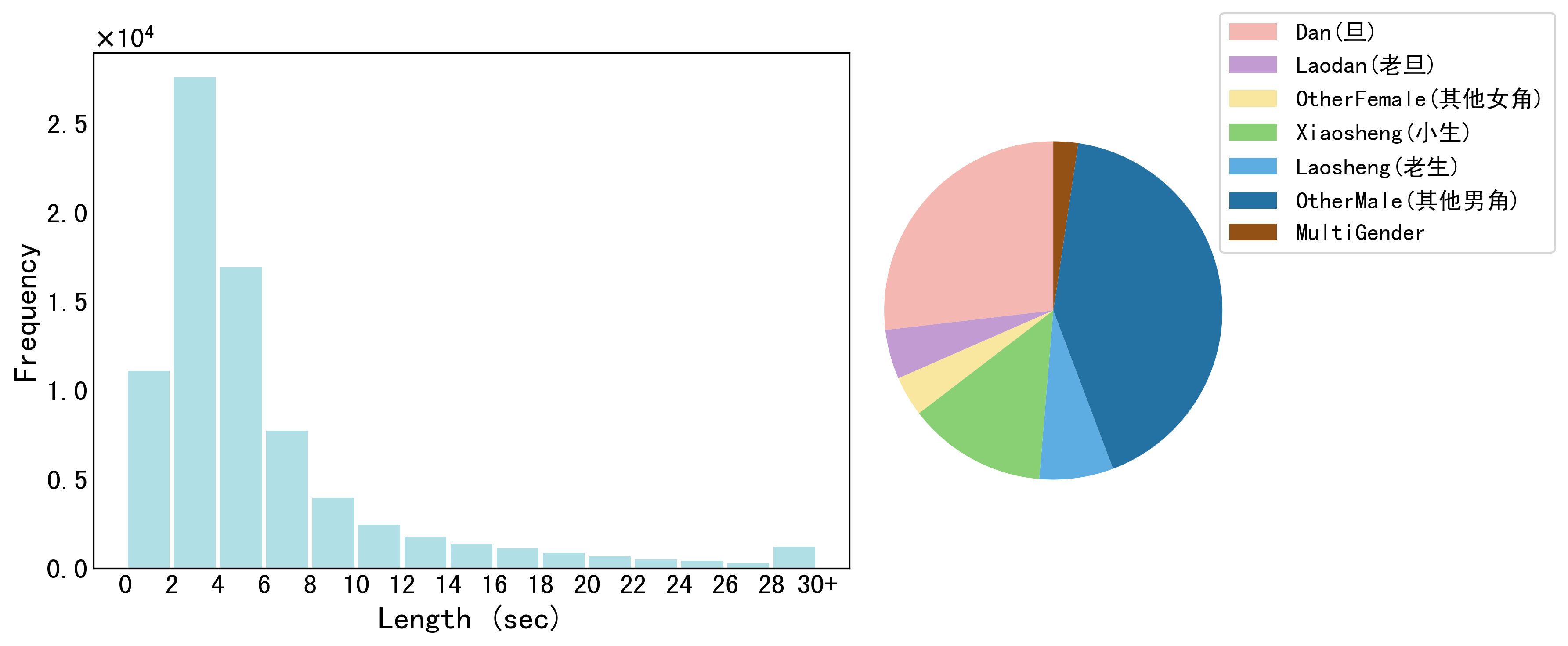}
  \caption{\textbf{Left:} Histogram of utterance lengths in the dataset. \textbf{Right:} Distribution of speaker role type information. The legend indicates the role type performed by speakers throughout the dataset. \textbf{\textit{Dan}} for young female characters, \textbf{\textit{LaoDan}} for old female characters, \textbf{\textit{OtherFemale}} for additional female characters; \textbf{\textit{XiaoSheng}} for young male characters, \textbf{\textit{LaoSheng}} for old male characters, \textbf{\textit{OtherMale}} for additional male characters; and \textbf{\textit{MultiGender}} means speakers portraying characters of both genders.}
  \label{fig-duration_role_dist}
\end{figure}

\vspace{-5pt}
Table~\ref{tab-dataset_stats} summarizes key statistics for the KunquDB dataset, differentiating stage speech (\textbf{\textit{ST}}) and singing (\textbf{\textit{S}}) categories. The dataset contains $60,066$ \textbf{\textit{ST}} utterances and $17,902$ \textbf{\textit{S}} utterances, contributed by $288$ speakers for both \textbf{\textit{ST}} and \textbf{\textit{S}} data, $50$ exclusively providing \textbf{\textit{ST}} data, and $1$ exclusively offering \textbf{\textit{S}} data. Additionally, there are $339$ videos featuring both \textbf{\textit{ST}} and \textbf{\textit{S}}, along with $5$ exclusively for \textbf{\textit{ST}} and $2$ for \textbf{\textit{S}}. Fig.~\ref{fig-duration_role_dist} visually represents the distribution of utterance lengths and speakers enacting role types.
\begin{table}[t]\footnotesize
  \caption{Dataset statistics for KunquDB}
  \label{tab-dataset_stats}
  \centering
  \vspace{-0.6em}
  \begin{tabular}{l c c }
    \toprule
    \textbf{Types of Utterances} & \textbf{Stage Speech} & \textbf{Singing} \\
    \midrule
    \# of speakers & $288+50$  & $288+1$~~~ \\
    \# of videos & $339+5$  & $339+2$~~~ \\
    \# of utterances & $60066$  & $17902$~~~ \\
    \# of hours & $67.46$  & $60.88$~~~ \\
    Avg \# of videos per speaker & $3$  & $3$~~~ \\
    Avg \# of utterances per speaker & $178$  & $62$ \\
    Avg length of utterances(s) & $4.04$  & $12.24$~~~ \\
    \bottomrule
  \end{tabular}
\end{table}

\vspace{-10pt}
\subsection{Split: Training and Test}
\vspace{-5pt}
We divide speakers based on their total number of utterances, with the initial $200$ individuals allocated to the training set and the remaining $139$ to the test set. See Table~\ref{tab-train_test_split} for details.

\begin{table}[t]\footnotesize
  \caption{Training and test data split}
  \label{tab-train_test_split}
  \centering
  \vspace{-0.6em}
  \begin{tabular}{lccc}
    \toprule
    & & \textbf{\#Speakers} & \textbf{\#Utterances} \\
    \midrule
    \multirow{2}{*}{\textbf{Training}} & Stage Speech & $200$ & $55889$\\ \cmidrule{2-4}
    & Singing & $200$ & $16941$\\ 
    \midrule
    \multirow{2}{*}{\textbf{Test}} & Stage Speech & $88+50$ & $4177$\\ \cmidrule{2-4}
    & Singing & $88+1$ & $961$\\ 
    \bottomrule
  \end{tabular}
\end{table}

\subsection{Trial Construction}
\vspace{-5pt}
When generating test trials for speaker verification experiments, we adopt a consistent procedure for each utterance, randomly selecting five positive and five negative samples. Investigating four trial scenarios considering two vocal manners (stage speech and singing), we have:

\begin{asparaitem}
    \label{four-trials}
    \item Undifferentiated Trial: No distinction between enrollment and test utterance regarding vocal manner categories; samples are randomly chosen from either stage speech or singing.
    \item Stage Speech Domain Trial: Both enrollment and test utterances are from the stage speech category.
    \item Singing Domain Trial: Both enrollment and test utterances are from the singing category.
    \item Cross-domain Trial: Enrollment is from singing, while test utterances are from the stage speech category.
\end{asparaitem}

\vspace{-5pt}
\section{Learning Domain-invariant Speaker Embeddings}
\subsection{Domain Discrepancy Adversarial Learning}\label{subsec-ddal}
\begin{figure}[t]\footnotesize
  \centering
  \includegraphics[width=\linewidth]{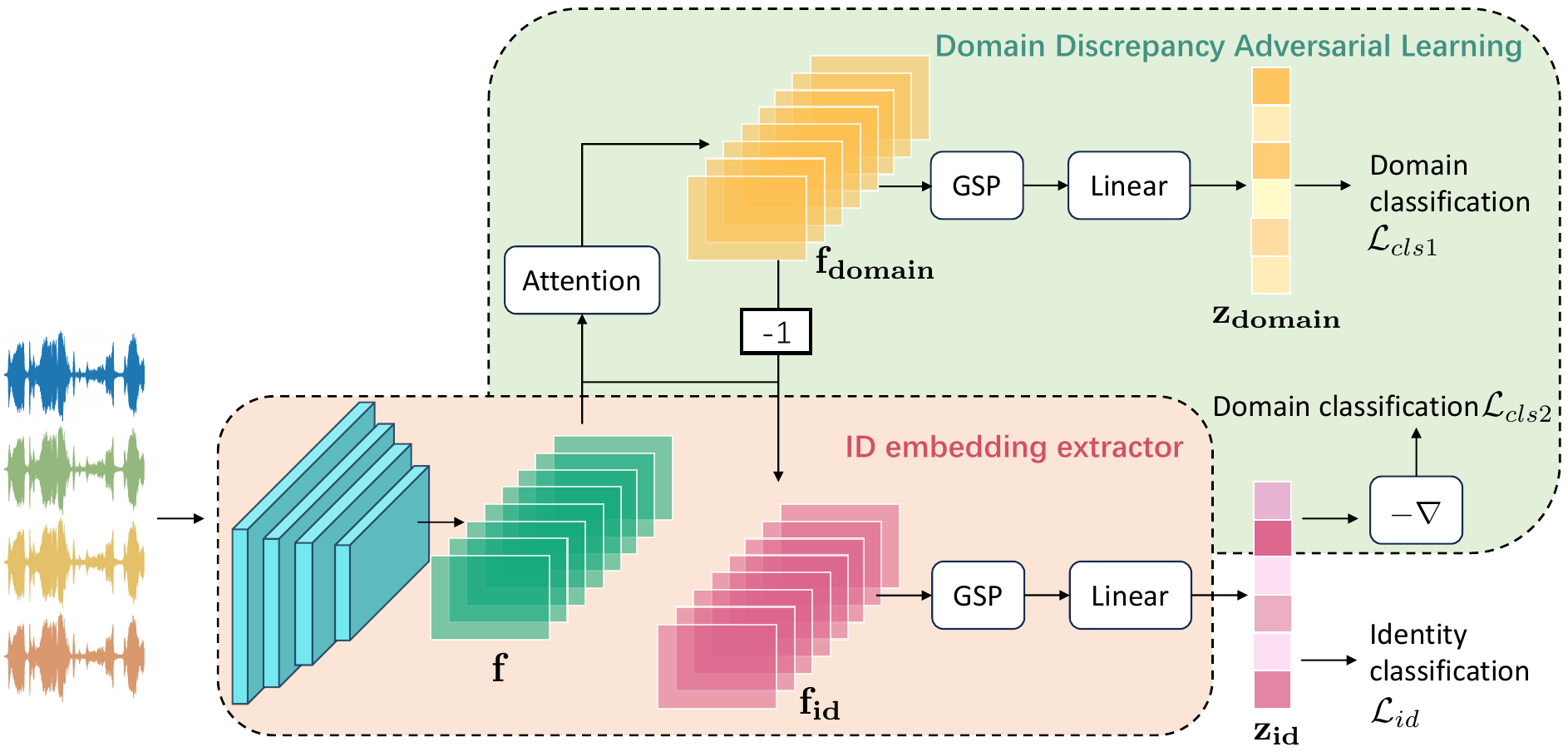}
  \caption{Schematic of the DDAL framework. The pink dashed box outlines the identity embedding extractor; the green dashed box highlights the core components of the DDAL mechanism.}
  \label{fig-ddal}
\end{figure}

\vspace{-5pt}
As discussed in Section~\ref{subsec-asv}, the speaker ID embedding extractor comprises a feature encoder, pooling, and linear layer. Traditionally, it is assumed that this extractor, depicted by the pink dashed box in Fig.~\ref{fig-ddal}, exclusively captures acoustic features defining speaker identity, denoted by the equation \(\mathbf{f} = \mathbf{f_{id}}\). However, it may inadvertently conflate identity-specific traits with variations from intrinsic factors like vocal mannerisms, formalized as Equation~\ref{equation-f_fid_fdomain}, where \(\mathbf{f}\) denotes the extracted features, \(\mathbf{f_{id}}\) refers to the identity-specific features, and \(\mathbf{f_{domain}}\) represents features associated with vocal manners.
\begin{equation}
    \label{equation-f_fid_fdomain}
    \setlength{\abovedisplayskip}{3pt}
    \setlength{\belowdisplayskip}{3pt}
    \mathbf{f}=\mathbf{f_{id}}+\mathbf{f_{domain}}
\end{equation}

Borrowing insights from \cite{qin2022cross}, we implement an optimized multi-task paradigm called Domain Discrepancy Adversarial Learning (DDAL), as illustrated in Fig.~\ref{fig-ddal}, to isolate domain-specific variables from speaker embeddings. This framework integrates speaker identity verification, domain classification, and domain adversarial training. Diverging from \cite{qin2022cross}, we disentangle domain characteristics at the feature map layer instead of the abstract embedding space. This early disentanglement capitalizes on the richer domain-specific details in the feature map layer, facilitating a cleaner separation and enhancing verification precision across domains.

We leverage an attention mechanism to disentangle domain-related features \(\mathbf{f_{domain}}\) induced by different vocal manners from the feature map \(\mathbf{f}\) extracted by the backbone model. Next, we refine speaker-specific features, \(\mathbf{f_{id}}\), by filtering out \(\mathbf{f_{domain}}\). Following this, both \(\mathbf{f_{domain}}\) and \(\mathbf{f_{id}}\) undergo pooling and fully connected layers, producing the domain embedding \(\mathbf{z_{domain}}\) for domain classification and speaker ID embedding \(\mathbf{z_{id}}\) for speaker classification. Further, we employ a gradient reversal layer (GRL) before an auxiliary domain classifier to eliminate domain influence from \(\mathbf{z_{id}}\) through adversarial learning.

Equation~\ref{equation-loss_ddal} defines the composite loss function, comprising the standard identity loss \(\mathcal{L}_{id}\) and the weighted sum of domain classifier losses \(\mathcal{L}_{cls1}\) and \(\mathcal{L}_{cls2}\). The weight \(\lambda_{ddal}\) acts as a tuning hyperparameter to balance these components:
\begin{equation}
\label{equation-loss_ddal}
    \mathcal{L}_{DDAL}=\mathcal{L}_{id}+\lambda_{ddal}(\mathcal{L}_{cls1}+\mathcal{L}_{cls2})
\end{equation}

\vspace{-5pt}
\subsection{Batchwise Contrastive Siamese Training}\label{subsec-bcst}
\begin{figure}[t]\footnotesize
  \centering
  \includegraphics[width=\linewidth]{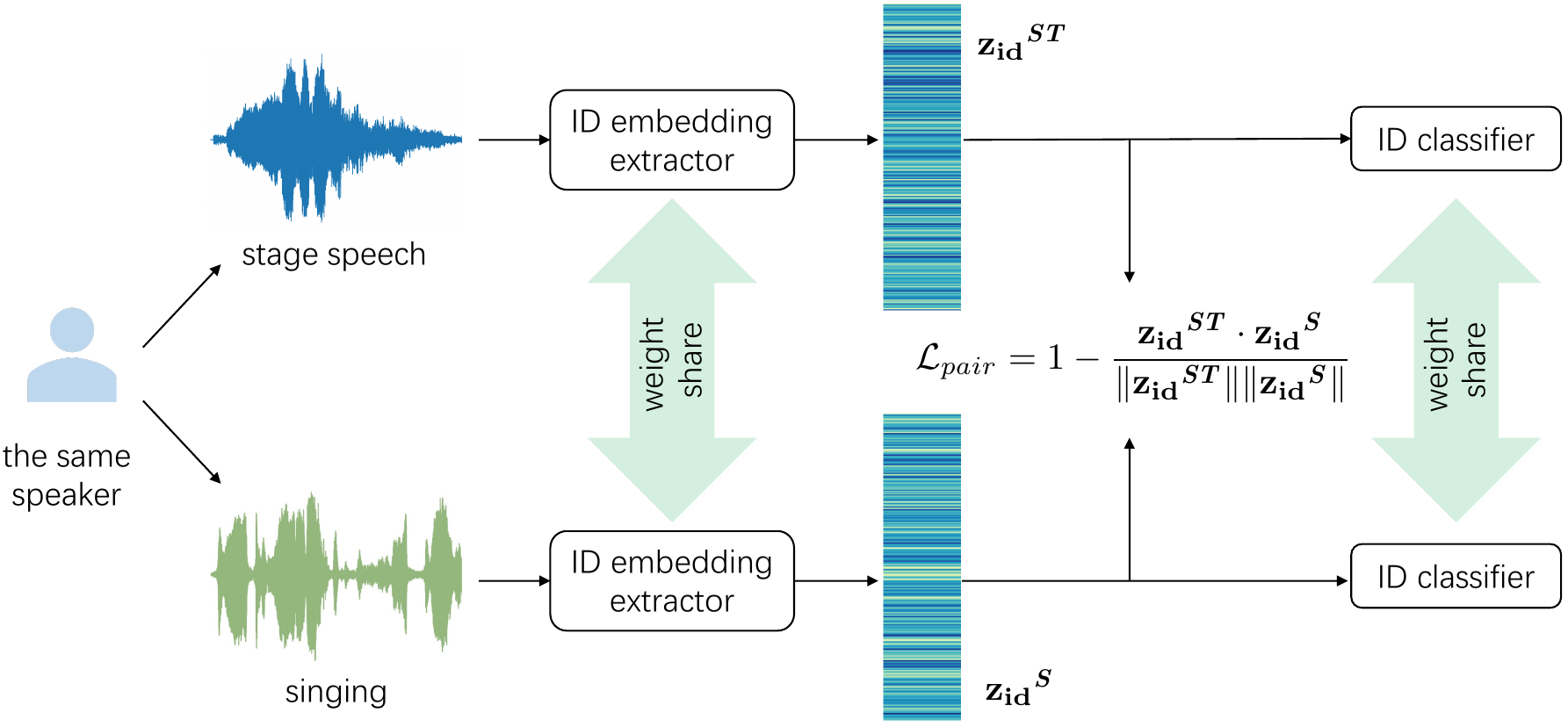}
  \caption{Overview of the BCST structure}
  \label{fig-siamese}
\end{figure}

\vspace{-5pt}
To effectively utilize utterances from the same speakers, we adopt a Batchwise Contrastive Siamese Training (BCST) strategy, inspired by \cite{lin2023haha}, to refine speaker embeddings across different domains into a unified, domain-independent representation. As depicted in Fig.~\ref{fig-siamese}, the model receives paired utterances from the same speaker but in different vocal manners.

The optimization process focuses on the $\mathcal{L}_{BCST}$, a combined loss comprising the individual utterance losses, $\mathcal{L}_{utt\textbf{\textit{S}}}$ and $\mathcal{L}_{utt\textbf{\textit{ST}}}$, as well as the pair loss $\mathcal{L}_{pair}$ scaled by a factor $\lambda_{bcst}$. The pair loss quantifies the cosine distance between speaker embeddings, \(\mathbf{z_{id}}^{\textbf{\textit{ST}}}\) and \(\mathbf{z_{id}}^{\textbf{\textit{S}}}\), extracted from paired utterances. By leveraging both singular utterance traits and relational information from utterance pairs, the model is encouraged to enhance its ability to distinguish between speakers and maintain feature consistency for the same speaker, even when their vocal manner varies.
\begin{equation}
\label{equation-loss_bcst}
    \setlength{\abovedisplayskip}{4pt}
    \setlength{\belowdisplayskip}{2pt}
    \mathcal{L}_{BCST} = \mathcal{L}_{utt\textbf{\textit{S}}}+\mathcal{L}_{utt\textbf{\textit{ST}}}+\lambda_{bcst}\mathcal{L}_{pair}
\end{equation}

\begin{equation}
\label{equation-loss_pair}
    \setlength{\abovedisplayskip}{2pt}
    \setlength{\belowdisplayskip}{3pt}
    \mathcal{L}_{pair} = 1-\frac{\mathbf{z_{id}}^{\textbf{\textit{ST}}} \cdot \mathbf{z_{id}}^{\textbf{\textit{S}}}}{\|\mathbf{z_{id}}^{\textbf{\textit{ST}}}\|\|\mathbf{z_{id}}^{\textbf{\textit{S}}}\|}
\end{equation}

\vspace{-10pt}
\section{Experiments}
\subsection{Experimental Setup}
\subsubsection{Dataset} 
We pretrain the model on VoxBlink2~\cite{lin2024voxblink2} with over 16,000 hours of audio data from 110k speakers. Thereupon, we fine-tune the model using KunquDB's training set. Evaluation is performed on the KunquDB test set.

\vspace{-10pt}
\subsubsection{Network} 
In our baseline (detailed in Table~\ref{tab-renet34_arch}), we use ResNet34~\cite{he2016deep} as the feature extractor, followed by a Global Statistic Pooling (GSP) layer to condense the length-variable frame-level feature map into a fixed-length representation. This representation is then input to a fully connected layer with $256$ dimensions. For speaker identification, we employ the ArcFace classifier~\cite{deng2019arcface} (m=0.2, s=32). Binary domain classifier involves stacking Linear-ReLU-Linear structures on \(\mathbf{z_{domain}}\) and \(\mathbf{z_{id}}\) for domain classification and adversarial learning, respectively. In the attention mechanisms that decouple domain-related features \(\mathbf{f_{domain}}\) from global features \(\mathbf{f}\), we employ two approaches: a neural network-based method known as Attentive Statistics Pooling (ASP)~\cite{okabe2018attentive} and a Simple, Parameter-free Attention Module (SimAM)~\cite{yang2021simam}.
\begin{table}[t]\footnotesize
  \renewcommand\arraystretch{0.75}
  \caption{The architecture of our ResNet34 backbone network. The residual building blocks are shown in $\left[\cdot \right]$, with the numbers of blocks stacked. Downsampling is performed by Layer2\_1, Layer3\_1, Layer4\_1 with a stride of 2.}
  \label{tab-renet34_arch}
  \centering
  \vspace{-0.7em}
  \begin{tabular}[c]{l l c}
    \toprule
    \textbf{Layer} & \textbf{Structure} & \textbf{Output Size} \\
    \midrule
    Conv1 & $3\times 3, 64$ & $64 \times 80 \times T$ \\
    \midrule
        Layer1 
        & $\begin{bmatrix}
            3\times 3, 64 \\
            3\times 3, 64
        \end{bmatrix} \times 3$
        & $64 \times 80 \times T$ \\
    \midrule
        Layer2 
        & $\begin{bmatrix}
            3\times 3, 128 \\
            3\times 3, 128
        \end{bmatrix}\times 4$
        & $128 \times 40 \times \dfrac{T}{2}$ \\
    \midrule
        Layer3 
        & $\begin{bmatrix}
            3\times 3, 256 \\
            3\times 3, 256
        \end{bmatrix}\times 6$
        & $256 \times 20 \times \dfrac{T}{4}$ \\
    \midrule
        Layer4 
        & $\begin{bmatrix}
            3\times 3, 512 \\
            3\times 3, 512
        \end{bmatrix}\times 3$
        & $512 \times 10 \times \dfrac{T}{8}$ \\
    \midrule
    Encoding & Global Statistics Pooling & $1024$ \\
    \midrule
    ID Embedding & Linear & $256$ \\
    Domain Embedding & Linear & $256$ \\
    \bottomrule
  \end{tabular}
\end{table}

We initialize the baseline model by pre-training on the VoxBlink2 dataset and experiment with various fine-tuning strategies using the KunquDB training set, as detailed in Table~\ref{tab-models}. \textbf{M0} serves as the standard and starting point for all subsequent fine-tuning experiments; it is pre-trained but not fine-tuned. \textbf{M1} and \textbf{M2} undergo fine-tuning using the standard ResNet34-GSP architecture, aligning with \textbf{M0}. In contrast, \textbf{M3} and \textbf{M4} are built on the SimAM-based DDAL framework; likewise, \textbf{M5} and \textbf{M6} adopt the ASP-based DDAL approach. \textbf{M2}, \textbf{M4}, and \textbf{M6} incorporate the BCST strategy, further building upon \textbf{M1}, \textbf{M3}, and \textbf{M5}, respectively.

\begin{table}[t]\footnotesize
  \renewcommand\arraystretch{0.9}
  \caption{Models varied in architectures, training data, and strategies. \textbf{KunquDB fine-tuning} indicates whether to utilize the KunquDB training set for fine-tuning. \textbf{DDAL} denotes Domain Discrepancy Adversarial Learning as described in Section~\ref{subsec-ddal}; \textbf{BCST} refers to Batchwise Contrastive Siamese Training as detailed in Section~\ref{subsec-bcst}.}
  \label{tab-models}
  \centering
  \vspace{-0.7em}
  \begin{tabular}{@{}lcccc@{}}
    \toprule
        \multirow{2}*{\textbf{ID}} & \multirow{2}*{\textbf{Model}} & \multirow{2}*{\textbf{Size}} & \textbf{KunquDB} & \multirow{2}*{\textbf{BCST}} \\
         & & & \textbf{fine-tuning} & \\
    \midrule
        \textbf{M0} & \multicolumn{1}{l}{\multirow{3}*{\quad ResNet34-GSP}} & \multirow{3}*{20.54M} & \texttimes & \texttimes \\
        \textbf{M1} & ~ & ~ & \checkmark & \texttimes \\
        \textbf{M2} & ~ & ~ & \checkmark & \checkmark \\
    \midrule
        \textbf{M3} & \multicolumn{1}{l}{\multirow{2}*{\quad\quad + SimAM-based DDAL}} & \multirow{2}*{20.79M} & \checkmark & \texttimes \\
        \textbf{M4} &  & ~ & \checkmark & \checkmark \\
    \midrule
        \textbf{M5} & \multicolumn{1}{l}{\multirow{2}*{\quad\quad + ASP-based DDAL}} & \multirow{2}*{27.35M} & \checkmark & \texttimes \\
        \textbf{M6} & ~ & ~ & \checkmark & \checkmark \\
    \bottomrule
  \end{tabular}
\end{table}

\vspace{-10pt}
\subsubsection{Training Details} 
During pre-training, we apply on-the-fly data augmentation~\cite{cai2020fly} and follow a training setting similar to \cite{qin2022dku}. For fine-tuning, we utilize a multi-step learning rate (LR) scheduler starting with an initial LR of $10^{-3}$ to modulate the SGD optimizer, gradually updating the model parameters until convergence. The hyperparameters $\lambda_{ddal}$ and $\lambda_{bcst}$ are assigned with a value of $0.5$ when used independently within the model (\textbf{M1, M2, M3, M5}). However, when both are employed (\textbf{M4, M6}), $\lambda_{ddal}$ is set to $1$, while $\lambda_{bcst}$ is adjusted to $1.5$. Input utterances are truncated to 2 seconds and converted to 80-dimensional log Mel-ﬁlterbank energies.

\vspace{-10pt}
\subsubsection{Evaluation Metrics} 
Cosine similarity is used for trial scoring. The verification performances are measured by the Equal Error Rate (EER) and the minimum normalized detection cost function (mDCF) with $P_{target}=0.01$.

\vspace{-10pt}
\subsubsection{Experimental Results}
Table~\ref{tab-eer_results} reports the performance of models on different test sets, with several key observations discussed below.
\begin{table}[t]\footnotesize
    \caption{The performance comparison of different speaker verification systems in terms of Equal Error Rate (EER) across four distinct test sets, as outlined in Section~\ref{four-trials}.
    }
    \label{tab-eer_results}
    \centering
    \vspace{-0.6em}
    \begin{tabular}{@{}lcccccccc@{}}
    \toprule
        \multirow{2}*{\textbf{ID}} & \multicolumn{2}{c}{\textbf{Undifferentiated}} & \multicolumn{2}{c}{\textbf{\textbf{\textit{ST}}-Domain}}  & \multicolumn{2}{c}{\textbf{\textbf{\textit{S}}-Domain}} & \multicolumn{2}{c}{\textbf{Cross-Domain}} \\
        \cmidrule(lr){2-3} \cmidrule(lr){4-5} \cmidrule(lr){6-7} \cmidrule(lr){8-9}
        ~ & \textbf{EER[\%]} & \textbf{mDCF} & \textbf{EER[\%]} & \textbf{mDCF} & \textbf{EER[\%]} & \textbf{mDCF} & \textbf{EER[\%]} & \textbf{mDCF} \\
    \midrule
    \textbf{M0} & 21.48 & 0.99 & 18.81 & 0.97 & 23.06 & 0.97 & 28.52 & 1.00 \\
    \textbf{M1} & 7.95  & 0.66 & 7.53 & 0.61 & 7.29  & 0.77 & 9.84  & 0.84 \\
    \textbf{M2} & 7.79  & 0.67 & 7.67 & 0.65 & 6.47  & 0.70 & 9.37  & 0.79 \\
    \textbf{M3} & 7.79  & 0.71 & 7.57 & 0.64 & 7.20 & 0.87 & 9.40 & 0.88 \\
    \textbf{M4} & \textbf{7.36} & 0.71 & \textbf{7.12} & 0.70 & \textbf{6.21} & 0.72 & 8.37 & 0.84 \\
    \textbf{M5} & 7.64 & 0.71 & 7.56 & 0.63 & 6.41 & 0.78 & 8.79 & 0.88 \\
    \textbf{M6} & 7.39 & 0.69 & 7.41 & 0.63 & 6.32 & 0.71 & \textbf{8.25} & 0.78 \\
    \bottomrule
\end{tabular}
\end{table}

\begin{asparaenum}[(1)]
\item Model \textbf{M0} shows weak robustness on Kunqu data, performing best in the \textbf{\textit{ST}}-domain due to its exclusive pretraining on speech data. Nevertheless, its performance is still markedly inferior to its excellent performance on regular speech test sets, often below $1\%$ EER.
\item Models generally perform best when enrollment and test utterances share the same vocal manner, whether in the \textbf{\textit{S}} or \textbf{\textit{ST}} category. However, their performance notably declines in cross-domain scenarios, indicating the difficulty in extracting domain-agnostic speaker embeddings.
\item DDAL or BCST individually improves model performance on Kunqu datasets. Deploying both approaches concurrently (\textbf{M4} and \textbf{M6}) substantially augments this enhancement, delivering superior outcomes.
\item Regarding the two implementations of attention within the DDAL strategy, the ASP-based implementation (\textbf{M5}) outperforms the SimAM-based counterpart (\textbf{M3}) across all test sets without BCST. However, with BCST integration, the SimAM-based approach (\textbf{M4}) yields better results than the ASP-based method (\textbf{M6}) in three out of four test sets, except for the cross-domain scenario.
\end{asparaenum}

\begin{figure}[t]\footnotesize
  \centering
  \includegraphics[width=\linewidth]{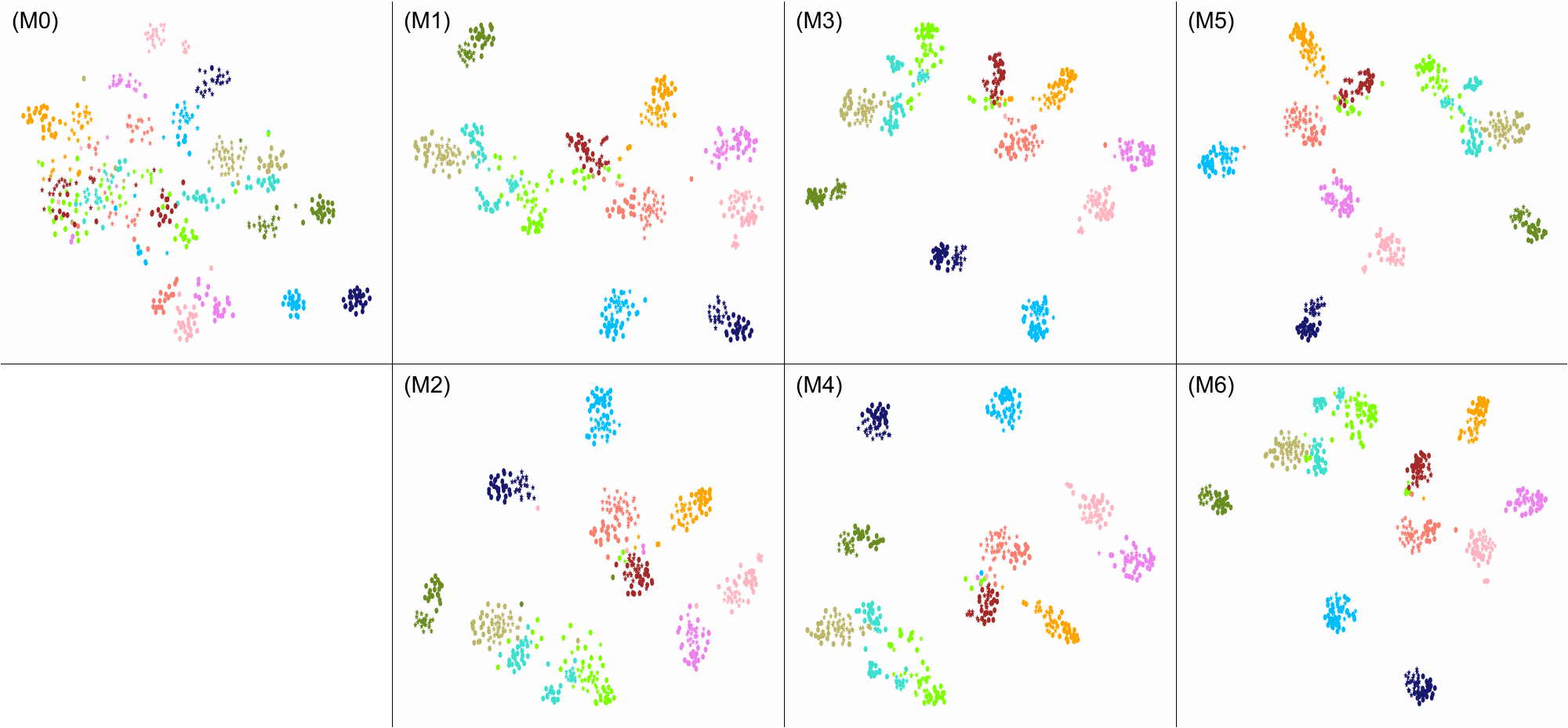}
  \caption{t-SNE visualization of speaker embedding extracted by seven models (\textbf{M0}$\sim$\textbf{M6}). Unique colors signify individual distinctions, with circular markers (\textbf{$\medbullet$}) representing stage speech utterances and pentagonal stars (\textbf{$\bigstar$}) denoting singing utterances.}
  \label{fig-tsne}
\end{figure}

We randomly select eleven individuals from the test data and visualize their speaker embeddings using the t-distributed stochastic neighbor embedding (t-SNE) algorithm in Fig.~\ref{fig-tsne}. Each subfigure corresponds to a specific model (\textbf{M0}$\sim$\textbf{M6}), providing a visual representation of the distribution patterns learned under various domain adaptation approaches. Notably, the \textbf{M0} subfigure reveals a lack of convergence in the distributions of utterances from the same speaker across different domains. In contrast, coherent distributions are observed among similar utterance types, with \textbf{\textit{S}} utterances predominantly in the left upper quadrant and \textbf{\textit{ST}} utterances in the right lower quadrant. These t-SNE visualizations consistently mirror the objective performance metrics presented in Table~\ref{tab-eer_results}, confirming the effectiveness of the domain adaptation methods.

\section{Conclusion}
This paper introduces KunquDB, a relatively large-scale, publicly accessible audio-visual dataset designed to address research gaps in Chinese opera studies. With detailed annotations, KunquDB aims to serve as a valuable resource for opera and speech-related research endeavors. Leveraging domain discrepancy adversarial learning and batchwise contrastive Siamese training, we establish benchmarks for ASV on Chinese opera data, offering unique insights distinct from conventional speech datasets.

\section{Acknowledgements}
This research is funded by the Kunshan Municipal Government Research Funding under the project "Deep Learning based Singing Voice Synthesis for Kun Opera". We want to thank the publisher for allowing us to conduct research on their data and DKU library staff members for their coordination. Special thanks to Xiaoyi Qin for his assistance.

\bibliographystyle{splncs04}
\bibliography{mybib}
\end{CJK}
\end{document}